\documentclass[traditabstract]{aa} 

\usepackage{graphicx, times, txfonts, float, rotating, color, url, lscape, bm} 
\usepackage{placeins,afterpage}

\newcommand{\less}{\raisebox{-1.1mm}{$\stackrel{<}{\sim}$}}

\newcommand{\ks}{km s$^{-1}$} 
 
\newcommand{\mum}{$\mu$m}
\newcommand{\muas}{$\mu$as} 

\newcommand{\G}{{\it Gaia}}
\newcommand{\Hp}{{\it Hipparcos}}

\begin{document}

\title{Orbital parallax of binary systems compared to GAIA DR3 and the parallax zero-point offset at bright magnitudes
\thanks{
Tables~\ref{Tab:Sam} and \ref{Tab-Par} are available in electronic form at the CDS via 
anonymous ftp to cdsarc.u-strasbg.fr (130.79.128.5) or via 
http://cdsweb.u-strasbg.fr/cgi-bin/qcat?J/A+A/. 
}
}  
 
\author{ 
M.~A.~T.~Groenewegen
}

\institute{ 
Koninklijke Sterrenwacht van Belgi\"e, Ringlaan 3, B--1180 Brussels, Belgium \\ \email{martin.groenewegen@oma.be}
} 
 
\date{received: ** 2022, accepted: 14 October 2022} 
 
\offprints{Martin Groenewegen} 
 
\titlerunning{The orbital parallax of binary systems compared to GAIA DR3} 
 
\abstract
{
Multiple systems for which the astrometric and spectroscopic orbit are known offer the unique possibility
of determining the distance to these systems directly without any assumptions. They are therefore ideal objects
for a comparison of  {\it Gaia} data release 3 (GDR3) parallax data, especially since GDR3 presents the
results of the non-single star (NSS) analysis that potentially results in improved parallaxes.
This analysis is relevant in studying the parallax zero-point offset (PZPO) that is crucial in improving upon the distance scale. \\
An sample of 192 orbital parallax determinations for 186 systems is compiled from the literature.
The stars are also potentially in wide binary systems (WBS). A search was performed and  37 WBS (candidates) were found. \\
Only for 21 objects does the NSS analysis provide information, including 8 from the astrometric binary pipeline, for which
the parallaxes do improve significantly compared to those in the main catalogue with significant lower goodness-of-fit (GOF) parameters.
  It appears that most of the objects in the sample are eliminated in the pre-filtering stage of the NSS analysis. \\
  The difference between the orbital parallax and the (best) \G\ parallax was finally obtained for  170 objects.
  A raw comparison is meaningless, however, due to limitations in accuracy both in the orbital and in \G\ data.
  As many systems have been eliminated in the pre-filtering stage of the astrometric NSS pipeline, they remain in GDR3 with
  values for the GOF parameter in the range from several tens to several hundreds. When objects with large parallax errors or
  unrealistically large differences between the orbital and \G\ parallaxes are eliminated, and objects with a
  GOF $< 100$ or $<8$ are selected (the latter also with $G<10.5$~mag selected),
  samples of 68 and 20 stars remain. Parallax differences in magnitude bins and for the sample are presented.
  Three recipes from the literature that calculate the PZPO are tested. After these corrections are applied the remaining
  parallax differences are formally  consistent with zero within the error bar for all three recipes.
  In all cases, an uncertainty in these averages of about 10-15~\muas\ remains for these samples due to the small number statistics. \\
The proof of concept of using orbital parallaxes is shown to work, but the full potential is not reached as an
  improved parallax from the NSS analysis is available for only for eight systems. 
  In the final selection, the orbital parallax of 18 of 20 stars is known to better than 5\%, and the parallax determination for 6
  stars is better than from \G.
  In the full sample, 148 objects reach this precision in orbital parallax and therefore the full potential of using orbital parallaxes may hopefully be reached with GDR4. 
}


\maketitle

\section{Introduction}
\label{S-Int}
With the second data release (DR; \citealt{GDR2Sum}) of the {\it Gaia} mission \citep{GC2016a}, it was demonstrated that 
the mean parallax of quasi-stellar objects (QSOs) was not zero, but is slightly negative, $-0.029$ mas \citep{Lindegren18}.
This was confirmed in the third early DR (GEDR3; \citealt{GEDR3_Brown}), where the average and median
parallax zero-point offset (PZPO) of QSOs are $-21$ and $-17$~\muas, respectively (\citealt{GEDR3_LindegrenZP}; hereafter L21).
L21 presented a Python script to the community that returned the PZPO (without an error bar) as a function of
input parameters, namely ecliptic latitude ($\beta$),
$G$-band magnitude, the {\tt astrometric\_params\_solved} parameter, and either
the effective wavenumber of the source used in the astrometric solution ($\nu_{\rm eff}$, {\tt nu\_eff\_used\_in\_astrometry} for the
five-parameter solution {\tt astrometric\_params\_solved} = 31) or the
astrometrically estimated pseudo-colour of the source  ({\tt pseudocolour)} for the six-parameter
solution ({\tt astrometric\_params\_solved} = 95).
The module is defined in the range 6 $< G <$ 21~mag, 1.72 $>$ $\nu_{\rm eff}$ $>$ 1.24~\mum$^{-1}$, corresponding to about
0.15 $< (G_\text{BP} - G_\text{RP}) <$ 3.0~mag where $G, G_\text{BP}$, and $G_\text{RP}$ are the magnitudes in
the {\it Gaia} G, Bp, and Rp band, respectively.
The script is based on the parallax values of QSOs and wide-binary systems (WBS; see L21 for all details).

In parallel, after the publication of GDR2 numerous studies appeared that studied the PZPO offset using other classes of stars,
typically at brighter magnitudes than the QSOs.
Examples using GDR2 data are studies of classical cepheids (CCs; \citet{RiessGDR2}, \citet{Gr_GDR2}) that found a more negative value
($-0.046 \pm 0.013$ and $-0.049 \pm 0.018$~mas, respectively),
RR Lyrae ($\sim-0.056$~mas, \citealt{Muraveva18}, $-0.042 \pm 0.013$~mas, \citealt{Layden19}),
red clump stars \citep{ChanBovy19},
(detached) eclipsing binaries \citep{Stassun18,Graczyk19},
red giant asteroseismic parallaxes  \citep{Zinn18,Khan19}, and
stars with parallaxes from very long baseline interferometry \citep{Xu19}, amongst others.
With the advent of GEDR3 this line of research has continued \citep{Huang21, StassunTorres21, Ren21, Zinn21, RiessGEDR3, MaizA22}, 
including claims that the L21 procedure overcorrects the PZPO \citep{RiessGEDR3, Zinn21}, at least for brighter objects.
Alternative procedures for the L21 correction have been proposed by \citet{Gr21} (hereafter G21) and \citet{MaizA22} (hereafter MA22).
Clearly, the spatial, magnitude, and colour dependence of the PZPO remains of great interest.

One disadvantage of using alternative classes of objects is the intrinsic assumption is that the distances to those objects are known exactly, and
with an accuracy comparable to that of {\it Gaia}. However, with the exception of the QSOs that truly have a parallax of zero for all practical
purposes, this is based on direct or indirect assumptions or model dependences, such as reddening, a (linear) period-luminosity relation,
an absolute magnitude, or a surface-brightness colour relation.

Binaries for which the astrometric and spectroscopic orbits are known offer the possibility of deriving the orbital parallax free from
any assumption, based on Kepler laws. The only limitation in the accuracy of the distance is ultimately the quality of the data.
GDR3 \citep{GaiaDR3Vallenari22} offers the possibility of comparing {\it Gaia} trigonometric parallaxes
to orbital parallaxes. In previous releases, the  {\it Gaia} astrometric solution assumed single stars.
Therefore, the quality parameters of the astrometric solution, such as the 
the goodness of fit (GOF, {\tt astrometric\_gof\_al}) or the 
the renormalised unit weight error ({\tt RUWE}), were (very) poor for (close) binary systems.
In GDR3, non-single star (NSS) solutions are considered \citep{Halbwachs22,GaiaDR2Arenou22}, which means that for a subset
of stars, (improved) parallaxes are determined that binary motion into account.
Individual studies of astrometric and spectroscopic binary orbits also often derived the orbital parallax and compared it
to \Hp\ or \G\ data, but no real systematic differences can be identified as the PZPO is a small quantity
using mostly single objects (\citealt{Gallenne19OP} used four objects in the comparison of the orbital parallax and
GDR2 data).

The paper is structured as follows.
In Section~\ref{S-Sam}, the sample of binaries is introduced and confronted with G(E)DR3 data.
The impact and limitations of the DR3 NSS analysis are discussed in Sections~\ref{SS-NSS} and \ref{SS-compare},
and 
the PZPO is  discussed in Section~\ref{SS-PZPO}.
A brief discussion and summary conclude the paper.

\section{Sample}
\label{S-Sam}

Recently, \citet{Piccotti20} compiled a list of 69 SB2s with both visual and spectroscopic orbits, but
the emphasis of that paper was on the masses and ages of the systems. 
The sample selection followed that of \citet{Piccotti20} and was based on an extensive literature search using  the
Sixth Catalogue of Orbits of Visual Binary Stars (ORB6; \citealt{Hartkopf01}, starting from orbits with grades 1, 2,
and 3)\footnote{\url{http://www.astro.gsu.edu/wds/orb6/orb6frames.html}} and the
Ninth Catalogue of Spectroscopic Binary orbits (SB9; \citealt{Pourbaix04})\footnote{\url{https://sb9.astro.ulb.ac.be}}.
This initial correlation pointed to other literature that was then searched.
Articles by specific authors were searched through the ADS, and the ArXiv was monitored for relevant papers.
The literature search ended May 10, 2022.
Table~\ref{Tab:Sam} lists the adopted orbital elements and velocity amplitudes with references for a total of 186 systems.
For six binaries, several components have been resolved (WDS 03272$+$0944, 06024$+$0939, 06290$+$2013, 17247$+$3802, 20396$+$0458,
and 22388+4419), and therefore the table has 192 entries.
This almost triples the number of systems compared to \citet{Piccotti20}.
Although it was attempted to make this list complete, this cannot be guaranteed.
Obviously, GDR3 provides orbital elements and velocity amplitudes for selected systems, and the impact of this is
discussed in Sect.~\ref{SS-compare}.

The orbital parallax follows from the orbital elements as
\begin{equation}
\pi_{\rm o} = a / \left( (K_1 + K_2) \cdot P/29.7847 \cdot \sqrt{1 - e^2}/ \sin i \right)
\end{equation}
where $a$ is the semi-major axis in mas, $K_1$ and $K_2$ are the velocity semi-amplitudes of the
two components (in \ks), $P$ is the orbital period in years, $e$ is the eccentricity of the orbit, 
$i$ is the inclination in degrees, and $\pi_{\rm o}$ is the orbital parallax in mas.
The error in $\pi_{\rm o}$ is calculated through standard error propagation assuming all errors are independent.
The orbital parallaxes and errors are listed in Table~\ref{Tab-Par}.

No selection on the accuracy of period, major axis, inclination, or velocity amplitude is made to be included in the sample,
even if a large error implies a large error on the orbital parallax and likely a value that is not competitive
in accuracy with the \G\ value.
For objects for which no error bars on the orbital elements was published, an error of 1.3, 6, and 40\% on period, 2.5, 15, and 40\%)
on $a$, 5, 12, and 27\degr on $i$, and 0.01, 0.02, and 0.10 on $e$ was adopted for orbits of grade 1, 2, and 3 in ORB6, respectively.
If no error on the velocity amplitude was reported, an error of 5\% was adopted.
These cases are marked in Table~\ref{Tab:Sam} and are good targets for further observations to improve on the orbital parameters.

\begin{sidewaystable*}

\setlength{\tabcolsep}{1.0mm}
\caption{\label{Tab:Sam} Sample of stars (selected entries) } 
\begin{tabular}{crrccccccccrrrrlllll} \hline  \hline
 Name & HD    &  Period  &  $\sigma_{\rm P}$  & u &  $a$        & $\sigma_{\rm a}$ & $i$      &  $\sigma_{\rm i}$ & $e$      &  $\sigma_{\rm e}$ & $K_{\rm 1}$  & $\sigma_{\rm K_1}$ & $K_{\rm 2}$  & $\sigma_{\rm K_2}$ & Reference   \\
      &       &          &                  &       &  (\arcsec)  &     (\arcsec)  &  (\degr)  &    (\degr)      &          &                 &  (\ks)      &   (\ks)          &  (\ks)      &   (\ks)         &      \\ 
\hline
%
00084$+$2905 &    358 &    96.7015 &   0.0044    & d &    0.02400 & 0.00013    &  105.60 &  0.23    &  0.5348 & 0.0046    &    27.74 &   0.55    &    65.47 &   0.96    &  \citet{Pourbaix2000}                                         \\ 
00369$+$3343 &   3369 &   143.53 &   0.06    & d &    0.006690 & 0.000050    &  103.0 &  0.2    &  0.5420 & 0.0060    &    47.50 &   0.53    &   117.4 &   2.8    &  \citet{Pearce36}  \citet{Hummel95}                           \\ 
00373$-$2446 &   3443 &  9165.640 &  10.592    & d &    0.6750 & 0.0169*   &   78.6 &  5.0*   &  0.2352 & 0.0096    &     5.13 &   0.29    &     6.93 &   0.22   &  \citet{Pourbaix2000}                                         \\ 
00490$+$1656 &   4676 &    13.82462 &   0.00002    & d &    0.006527 & 0.000061    &   73.80 &  0.92    &  0.2376 & 0.0012    &    57.35 &   0.31    &    59.95 &   0.32    &  \citet{Boden1999}                                            \\ 
01028$+$3148 &   6118 &    81.12625 &   0.00027    & d &    0.005560 & 0.000040    &  143.4 &  1.3    &  0.8956 & 0.0020    &    53.2 &   1.9    &    59.6 &   1.6    &  \citet{Konacki04}                                            \\ 
01096$-$4616 &        &    24.59215 &   0.00002    & d &    0.001315 & 0.000005    &   91.32 &  0.39    &  0.1872 & 0.0001    &    51.166 &   0.008    &    49.118 &   0.007    &  \citet{Gallenne19OP}              \\ 
01108$+$6747 &   6840 &  2722.0 &   1.0    & d &    0.08300 & 0.01245*   &   52.0 & 12.0*   &  0.7442 & 0.0020    &    11.18 &   0.06    &    12.34 &   0.12    &  \citet{Griffin12}  \citet{Docobo18Inf}                       \\ 
01237$+$3743 &   8374 &    35.36836 &   0.00005    & d &    0.005050 & 0.000020    &  140.64 &  0.45    &  0.6476 & 0.0005    &    39.27 &   0.05    &    40.47 &   0.05    &  \citet{Lester20}                                             \\ 
01277$+$4524 &   8799 &   254.9003 &   0.1960    & d &    0.0380 & 0.0010    &   62.49 &  2.10    &  0.142 & 0.012    &    17.54 &   0.30    &    19.62 &   0.30    &  \citet{Farrington14}                                         \\ 
01321$+$1657 &   9312 &    36.51836 &   0.00068    & d &    0.0030 & 0.0012*   &   65.4 & 13.3    &  0.1429 & 0.0003    &    34.972 &   0.006    &    45.821 &   0.065    &  \citet{Wang15} \citet{Kiefer18}                              \\ 
01374$+$2510 &   9939 &    25.20896 &   0.00007    & d &    0.004944 & 0.000018    &   61.56 &  0.25    &  0.10166 & 0.00097    &    34.952 &   0.055    &    44.68 &   0.24    &  \citet{Boden06}                                              \\ 
01376$-$0924 &  10009 &    28.83 &   0.25    & y &    0.2920 & 0.0090    &   97.4 &  0.2    &  0.748 & 0.005    &     9.16 &   0.21    &    11.21 &   0.32    &  \citet{Tokovinin93}                                          \\ 
01379$-$8259 &  10800 &     1.74861 &   0.00047    & y &    0.07823 & 0.00047    &   47.60 &  0.48    &  0.1912 & 0.0018    &     9.10 &   0.03    &    17.91 &   0.05    &  \citet{Tokovinin16}                                          \\ 
01437$+$5041 &  10516 &   126.6982 &   0.0035    & d &    0.005890 & 0.000020    &   77.6 &  0.3    &  0.00000 & 0.1*   &    10.2 &   1.0    &    81.5 &   0.7    &  \citet{Mourard15}                                            \\ 
01546$+$2049 &  11636 &   106.99442 &   0.00069    & d &    0.03600 & 0.00016    &   47.50 &  0.54    &  0.8801 & 0.0008    &    37.10 &   1.86*   &    66.60 &   3.33*   &  \citet{Pourbaix2000}  \citet{Tokovinin16}                    \\ 
02057$-$2423 &  12889 &     2.582 &   0.003    & y &    0.0480 & 0.0030    &   75.9 &  4.0    &  0.769 & 0.015    &    18.52 &   0.50    &    19.10 &   0.50    &  \citet{Tokovinin2014}                                        \\ 
02128$-$0224 &  13612 &     0.25952 &   0.00001    & y &    0.01398 & 0.00075    &   25.6 &  7.8    &  0.6920 & 0.0027    &    19.01 &   0.04    &    19.74 &   0.43    &  \citet{Anguita22}                                            \\ 
02171$+$3413 &  13974 &    10.02000 &   0.00010    & d &    0.009800 & 0.000060    &  167.0 &  3.0    &  0.0200 & 0.0050    &    10.52 &   0.17    &    11.85 &   0.40    &  \citet{Hummel95} \citet{Pourbaix2000}                        \\ 
02211$+$4246 &        &   318.0 &  17.0    & y &    0.940 & 0.022    &  147.0 &  2.3   &  0.8020 & 0.0072    &     1.90 &   0.57    &    2.92 &   0.57    &  \citet{Pourbaix2000}                                         \\ 
02262$+$3428 &  15013 &  2533.0 &   5.0    & d &    0.0990 & 0.0149*   &   49.9 & 12.0*   &  0.300 & 0.005    &     7.61 &   0.06   &   7.86 &   0.07    &  \citet{Docobo18Inf196} \citet{Griffin18}                     \\ 
02278$+$0426 &  15285 &  9182.4 &  14.6    & d &    0.5429 & 0.0052    &   73.0 &  0.7    &  0.210 & 0.010    &     5.60 &   0.30   &   6.40 &   0.30    &  \citet{Agati15}                                              \\ 
02415$-$7128 &  17215 &   106.3 &   4.6    & y &    0.568 & 0.012    &  114.7 &  0.4    &  0.384 & 0.020    &     4.12 &   0.44    &     4.69 &   0.88    &  \citet{Tokovinin22}                                          \\ 
02422$+$4012 &  16739 &   330.9910 &   0.0044    & d &    0.05318 & 0.00015    &  128.17 &  0.14    &  0.6630 & 0.0021   &    20.912 &   0.089    &    23.730 &   0.087    &  \citet{Pourbaix2000}  \citet{Bagnuolo06}                     \\ 
02442$-$2530 &  17134 &     6.68917 &   0.00510    & y &    0.10020 & 0.00060    &   42.00 &  0.72    &  0.4999 & 0.0015    &     7.440 &   0.041    &     8.050 &   0.041    &  \citet{Tokovinin16a}                                         \\ 
02539$-$4436 &  18198 &    51.46 &   0.82   & y &    0.2967 & 0.0044    &   49.5 &  1.9    &  0.774 & 0.013    &     3.98 &   0.61    &    9.93 &   0.70    &  \citet{Tokovinin16}                                          \\ 
03025$-$1516 &  18955 &    43.32032 &   0.00013    & d &    0.005810 & 0.000034    &   92.24 &  0.18    &  0.7594 & 0.0010    &    54.31 &   0.29   &   60.45 &  0.34   &  \citet{Halbwachs16}                                          \\ 
03048$+$5330 &  18925 &    14.5930 &   0.0046    & y &    0.14390 & 0.00073    &   90.60 &  0.71    &  0.7860 & 0.0038    &    13.67 &  0.22   &  18.57 &  0.31    &  \citet{Pourbaix99} \citet{Pourbaix2000}                      \\ 
03082$+$4057 &  19356 &   680.168 &   0.046    & d &    0.09343 & 0.00011    &   83.66 &  0.03    &  0.2270 & 0.0020    &    12.00 &   0.40    &    31.60 &   1.20    &  \citet{Hill71}  \citet{Baron12}                              \\ 
03147$-$3533 &  20301 &    75.66691 &   0.00019    & d &    0.002990 & 0.000011    &   85.71 &  0.04    &  0.00000 & 0.00010   &   38.91 &  0.01   &   40.88 &   0.01   &  \citet{Gallenne18_TZ}                                        \\ 
03272$+$0944 &  21364 &   145.113 &   0.071    & d &    0.01593 & 0.00010    &   86.67 &  0.12    &  0.2101 & 0.0053    &  43.160 &   2.410    &    38.370 &   0.190    &  \citet{Nemravova16}                                          \\ 
03272$+$0944 &  21364 &     7.14664 &   0.00002    & d &    0.00189 & 0.00011    &   86.85 &  0.22    &  0.01 & 0.10*   &  87.79 &   0.25   &   93.02 &  0.23   &  \citet{Nemravova16}                                          \\ 
03396$+$1823 &  22694 &     0.826 &   0.003    & y &    0.02250 & 0.00080    &  128.31 &  4.98    &  0.577 & 0.004    &  14.16 &   1.44   &    15.39 &  1.57   &  \citet{Jao16}                                                \\ 
03400$+$6352 &  17126 &    11.42 &   0.08    & y &    0.1460 & 0.0020    &   70.0 & 20.0*   &  0.475 & 0.016   &   7.22 &   0.14   &     9.95 &   0.36   &  \citet{Tokovinin19}                                          \\ 
03566$+$5042 &  24546 &    30.43885 &   0.00002    & d &    0.00699 & 0.00006    &   56.76 &  0.45    &  0.6421 & 0.0006    &    52.24 &   0.06   &    53.15 &   0.06    &  \citet{Lester20}                                             \\ 
04107$-$0452 &  26441 &    20.6290 &   0.0077    & y &    0.16472 & 0.00066    &   67.94 &  0.18    &  0.8400 & 0.0014    &    11.61 &   0.04    &    12.41 &   0.06    &  \citet{Anguita22}                                            \\ 
04142$+$2813 & 283447 &    51.1003 &   0.0022    & d &    0.002837 & 0.000035    &   63.3 &  1.1    &  0.2710 & 0.0072    &    35.74 &   0.66    &    43.00 &   1.40    &  \citet{Torres12}                                             \\ 
04184$+$2135 &  27176 &    11.3642 &   0.0071    & y &    0.13537 & 0.00016    &  123.88 &  0.25    &  0.1540 & 0.0016    &     7.50 &   0.32    &     8.96 &   0.11    &  \citet{Anguita22}                                            \\ 
04209$+$1352 &  27483 &     3.05911 &   0.00001    & d &    0.00126 & 0.00005    &   45.1 &  1.7    &  0.0 & 0.1*   &    71.55 &   0.27    &    72.88 &   0.38    &  \citet{Konacki04} \citet{Griffin12Hyad}                      \\ 
04247$+$0442 &  27935 &   156.38052 &   0.00009    & d &    0.011338 & 0.000022    &  103.133 &  0.072    &  0.85128 & 0.00003    &    37.335 &   0.003   &   50.322 &   0.040    &  \citet{Halbwachs20}                                          \\ 
04256$+$1556 &  27991 &  2295.0 &   4.0    & d &    0.1000 & 0.0031    &  125.0 &  1.6    &  0.7155 & 0.0097    &    11.24 &   0.19    &    13.61 &   0.15    &  \citet{Pourbaix2000} \citet{Griffin12Hyad}                   \\ 
04268$+$1052 & 286820 &    19.77 &   0.27    & y &    0.201 & 0.010    &   63.0 &  2.2    &  0.649 & 0.040    &     6.80 &   0.34*   &     9.50 &   0.48*   &  \citet{Tokovinin21INF} \citet{Griffin12Hyad}                 \\ 
\hline
\end{tabular}
\tablefoot{
Column~1. The identifier used in this paper. It follows the WDS convention.
Column~2. HD number.   
Column~3-5. Pulsation period, error and unit (y = years, d = days, h = hours, m = minutes).
An asterisk in column~5, 7, 9, 11, 13, or 15 indicates that the error bar was assumed.
Column~6-7. Semi-major axis and error in arcsec.
Column~8-9. Inclination and error in degrees.
Column~10-11. Eccentricity and error.
Column~12-15. Semi-amplitude of the velocity with error, for both components.
Column~16. References.
In the electronic version the values for the orbital parameters are given at fixed formats
  which implies that the number of significant digits is sometimes too large.

}
  \end{sidewaystable*}

As a first step and as preparation for the DR3 release, the objects were identified in GEDR3 and also in the \Hp\ catalogue
(the new reduction version by \citealt{vanL08}). The later proved essential in many cases as 
the epoch 1991.25 \Hp\ coordinates could be transformed to the epoch 2016.0 of GEDR3/GDR3 to properly identify the correct object
that was achieved by comparing coordinates, parallax, and magnitude.
It was then trivial to identify the correct object with the release of DR3, and
Table~\ref{Tab-Par} gives the orbital parallax (based on the data in Table\ref{Tab:Sam} and Eq.~1), and some information
from \Hp\ and DR3 for the 192 objects. DR3 includes some parameters that were used in the pre-filtering stage of the astrometric binary
pipeline \citep{Halbwachs22}; they are discussed in detail below.

Inspection of Table~\ref{Tab:Sam} reveals that quite a number of objects (59, or about 30\%) are not listed in GDR3, or are listed
without parallax.
The former are six of the very brightest objects (all have $V$ \less\ 2.7~mag). 
The others only have two-parameter solutions ({\tt astrometric\_params\_solved = 3}).
However, these objects may potentially be in WBS.
To investigate this further, the objects were correlated with the catalogue of \citet{ElBadry21} (based on GEDR3 data).
Fourteen matches were found at distances  up to 57\arcsec.                                             
However, objects without a parallax or proper motion in GEDR3 are obviously missing from \citet{ElBadry21}.
In a second step, all sources within 1\arcmin\ were retrieved from GDR3 and a potential list of WBS was compiled
based on the orbital parallax and criteria on the parallax difference so that the search would retrieve all 14 matches
in \citet{ElBadry21}.
In a final step, the GDR3 and \Hp\ parallaxes and proper motions were inspected to make a final list
of likely WBS, keeping only stars with Bp and Rp photometry.
The information of these  37 sources is listed in Table~\ref{Tab:WBS}.

\begin{sidewaystable*}

\small
\setlength{\tabcolsep}{1.6mm}
  \caption{\label{Tab-Par} Parallax data (selected entries) } 
\begin{tabular}{crrrrrrrrrrcrrrrcccc}
  \hline  \hline
 Name &     $\pi_{\rm o}$  & HIP   & $\pi$       & GOF & $H_{\rm p}$   &  Source ID &  $\pi$       &  GOF & RUWE & $G$-mag & NSS & ipdfmp & ipdha & Nv & signiC$^\star$  \\
      &       (mas)       &       &   (mas)     &     &    (mag)     &           &  (mas)       &      &      &  (mag) &     &  (\%)  &          &      &     \\ 
\hline
%
00084$+$2905 &  33.019 $\pm$  0.428 &    677 &  33.62 $\pm$  0.35 & 11.77 &  2.04 &  & & & & & & & & &  \\ 
00369$+$3343 &   3.565 $\pm$  0.066 &   2912 &   5.45 $\pm$  0.31 &  5.82 &  4.31 &  364936228611467520 &   5.656 $\pm$  0.147 &    0.72 &  1.02 &  4.30 & 0  &  0  & 0.03  & 16  & 0.052  \\ 
00373$-$2446 &  67.002 $\pm$  2.608 &   2941 &  64.93 $\pm$  1.85 & 28.19 &  5.71 & 2347260998051944448 &                     &  103.56 &  0.00 &  5.95 & 0  & 99  & 0.24  &  3  & 131.180 \\ 
00490$+$1656 &  43.289 $\pm$  0.437 &   3810 &  42.64 $\pm$  0.27 &  1.30 &  5.18 & 2781872793183604096 &  42.751 $\pm$  0.112 &    0.21 &  1.01 &  4.93 & 2  &  0  & 0.03  & 14  & 1.523  \\ 
01028$+$3148 &   8.859 $\pm$  0.203 &   4889 &   7.52 $\pm$  0.26 &  1.51 &  5.49 &  314142055583243904 &   8.454 $\pm$  0.096 &   $-$1.00 &  0.94 &  5.49 & 0  &  0  & 0.02  & 14  & 0.580  \\ 
01096$-$4616 &   5.904 $\pm$  0.022 &   5438 &   3.50 $\pm$  1.04 & -0.45 &  8.74 & 4933562966514402176 &   5.907 $\pm$  0.019 &    4.86 &  1.15 &  8.43 & 2  &  0  & 0.05  & 26  & 0.474  \\ 
01108$+$6747 &  16.639 $\pm$  2.626 &   5531 &  18.30 $\pm$  0.53 &  3.75 &  6.67 &  526537396786294784 &  16.924 $\pm$  0.156 &  116.57 &  8.62 &  6.42 & 0  &  0  & 0.03  & 24  & 2.093  \\ 
01237$+$3743 &  16.213 $\pm$  0.066 &   6514 &  15.66 $\pm$  0.30 & -0.41 &  5.67 &  323159631479218560 &  15.762 $\pm$  0.127 &   -0.20 &  0.99 &  5.54 & 0  &  0  & 0.02  & 13  & 0.510  \\ 
01277$+$4524 &  39.106 $\pm$  1.121 &   6813 &  34.94 $\pm$  0.31 &  2.17 &  4.92 &  397372371388455424 &  34.733 $\pm$  0.134 &    5.59 &  1.33 &  4.69 & 0  & 14  & 0.03  & 14  & 3.179  \\ 
01321$+$1657 &  10.162 $\pm$  4.083 &   7143 &  17.32 $\pm$  0.50 &  0.80 &  6.94 & 2592750264855961472 &  17.730 $\pm$  0.060 &   39.54 &  2.66 &  6.54 & 0  &  0  & 0.04  & 15  & 0.072  \\ 
01374$+$2510 &  23.683 $\pm$  0.113 &   7564 &  22.55 $\pm$  0.56 &  1.40 &  7.15 &  292069737612587648 &  22.510 $\pm$  0.041 &   15.55 &  1.90 &  6.76 & 0  &  0  & 0.03  & 13  & 0.125  \\ 
01376$-$0924 &  22.128 $\pm$  0.818 &   7580 &  24.76 $\pm$  0.90 &  1.96 &  6.35 & 2464859126762143616 &  22.970 $\pm$  0.630 &  217.16 & 20.70 &  6.13 & 0  &  6  & 0.19  & 16  & 3.565  \\ 
01379$-$8259 &  37.116 $\pm$  0.237 &   7601 &  36.52 $\pm$  0.28 &  2.45 &  6.01 & 4618008180223988352 &  37.013 $\pm$  0.229 &   75.55 &  7.28 &  5.70 & 0  &  0  & 0.02  & 18  & 2.863  \\ 
01437$+$5041 &   5.387 $\pm$  0.078 &   8068 &   4.54 $\pm$  0.20 & -1.33 &  3.98 &  405987526029851264 &   5.405 $\pm$  0.197 &   13.53 &  1.80 &  4.08 & 0  &  0  & 0.08  & 15  & 6.557  \\ 
01546$+$2049 &  54.813 $\pm$  1.974 &   8903 &  55.60 $\pm$  0.58 & 28.01 &  2.70 &  & & & & & & & & &  \\ 
02057$-$2423 &  22.331 $\pm$  1.457 &   9774 &  21.44 $\pm$  1.43 &  2.07 &  8.70 & 5121561656518510080 &  21.089 $\pm$  0.088 &   39.98 &  3.00 &  8.41 & 0  & 72  & 0.16  & 17  & 23.693 \\ 
02128$-$0224 &  24.783 $\pm$  2.616 &  10305 &  25.19 $\pm$  1.41 & 25.87 &  5.77 & 2494066897240089728 &  26.538 $\pm$  0.127 &   28.82 &  3.28 &  5.54 & 0  &  0  & 0.02  & 11  & 1.491  \\ 
02171$+$3413 & 107.018 $\pm$  6.048 &  10644 &  92.73 $\pm$  0.39 &  5.00 &  4.98 &  326173044957518080 &  91.650 $\pm$  0.427 &   26.48 &  3.25 &  4.67 & 0  &  0  & 0.15  & 10  & 2.039  \\ 
02211$+$4246 &  16.655 $\pm$  2.660 &  10952 &  17.98 $\pm$  1.38 &  1.66 &  8.93 &  339384295642198272 &  15.417 $\pm$  0.236 &   35.90 &  4.20 &  9.19 & 0  & 52  & 0.10  & 10  & 150.764 \\ 
02262$+$3428 &  22.039 $\pm$  3.510 &  11352 &  24.31 $\pm$  0.99 &  1.67 &  8.15 &  326940164774368384 &  22.494 $\pm$  0.136 &   71.40 &  6.23 &  7.82 & 0  &  0  & 0.10  & 15  & 0.740  \\ 
02278$+$0426 &  52.428 $\pm$  1.879 &  11452 &  58.33 $\pm$  1.08 & -0.53 &  8.78 & 2515752565074318720 &                     & 1267.22 &  0.00 &  8.50 & 0  & 84  & 0.18  & 10  & 111.003 \\ 
02415$-$7128 &  17.775 $\pm$  2.001 &  12548 &  18.60 $\pm$  0.81 &  2.52 &  7.90 & 4644298430956871040 &                     &  574.00 &  0.00 &  7.63 & 0  & 36  & 0.21  & 24  & 6.940  \\ 
02422$+$4012 &  41.119 $\pm$  0.163 &  12623 &  41.34 $\pm$  0.43 &  8.42 &  5.02 &  335144892337326336 &  41.796 $\pm$  0.296 &   20.27 &  2.75 &  4.72 & 0  &  0  & 0.17  & 10  & 7.530  \\ 
02442$-$2530 &  22.253 $\pm$  0.158 &  12780 &  24.20 $\pm$  1.16 &  8.22 &  7.09 & 5076722404106123264 &  21.676 $\pm$  0.293 &  190.55 & 13.87 &  6.82 & 0  &  1  & 0.01  & 21  & 1.372  \\ 
02539$-$4436 &  14.826 $\pm$  0.998 &  13498 &  14.09 $\pm$  0.73 &  1.99 &  7.84 & 4755117967402573056 &  11.965 $\pm$  0.368 &  227.03 & 18.02 &  7.71 & 0  & 61  & 0.14  & 22  & 29.454 \\ 
03025$-$1516 &  19.526 $\pm$  0.137 &  14157 &  19.78 $\pm$  1.10 &  0.69 &  8.59 & 5154084145316239744 &  19.446 $\pm$  0.052 &   14.83 &  1.86 &  8.23 & 0  &  0  & 0.01  & 15  & 0.106  \\ 
03048$+$5330 &  14.735 $\pm$  0.188 &  14328 &  13.41 $\pm$  0.51 & 20.17 &  3.06 &  447071293401293056 &  14.125 $\pm$  0.768 &  134.43 & 10.85 &  2.65 & 0  &  1  & 0.06  & 18  & 47.001 \\ 
03082$+$4057 &  34.978 $\pm$  0.990 &  14576 &  36.27 $\pm$  1.40 & 52.82 &  2.10 &  & & & & & & & & &  \\ 
03147$-$3533 &   5.373 $\pm$  0.020 &  15092 &   5.75 $\pm$  0.51 &  2.80 &  7.02 & 5047189109469319680 &   5.471 $\pm$  0.019 &    9.10 &  1.31 &  6.68 & 2  &  0  & 0.01  & 27  & 0.738  \\ 
03272$+$0944 &  14.957 $\pm$  0.441 &  16083 &  15.60 $\pm$  1.04 & 31.49 &  3.71 &   12730974556132096 &  16.791 $\pm$  0.694 &   61.89 &  7.28 &  3.70 & 0  & 57  & 0.04  & 11  & 1.088  \\ 
03272$+$0944 &  15.889 $\pm$  0.929 &  16083 &  15.60 $\pm$  1.04 & 31.49 &  3.71 &   12730974556132096 &  16.791 $\pm$  0.694 &   61.89 &  7.28 &  3.70 & 0  & 57  & 0.04  & 11  & 1.088  \\ 
03396$+$1823 &  26.378 $\pm$  2.052 &  17076 &  25.61 $\pm$  1.34 &  1.84 &  8.39 &   56861694804053120 &  27.449 $\pm$  0.027 &    1.36 &  1.06 &  8.03 & 2  &  0  & 0.02  & 15  & 0.480  \\ 
03400$+$6352 &  23.682 $\pm$  1.927 &  17126 &  24.59 $\pm$  1.05 & -0.03 &  8.43 &  488099260551658752 &  24.067 $\pm$  0.573 &  472.07 & 28.84 &  8.06 & 0  &  4  & 0.06  & 22  & 4.136  \\ 
03566$+$5042 &  25.862 $\pm$  0.223 &  18453 &  26.71 $\pm$  0.87 & 17.82 &  5.38 &  250437313946591360 &  26.217 $\pm$  0.082 &    1.10 &  1.04 &  5.17 & 2  &  0  & 0.03  & 17  & 1.594  \\ 
04107$-$0452 &  16.912 $\pm$  0.085 &  19508 &  16.09 $\pm$  0.65 &  1.48 &  7.50 & 3203497397487659648 &  17.289 $\pm$  0.538 &  252.83 & 24.68 &  7.23 & 0  &  0  & 0.01  & 17  & 2.686  \\ 
04142$+$2813 &   7.119 $\pm$  0.163 &  19762 &  10.45 $\pm$  2.68 &  3.26 & 10.86 &  163184366130809984 &   8.326 $\pm$  0.131 &  119.48 &  7.53 &  9.99 & 0  &  0  & 0.01  & 18  & 1.568  \\ 
04184$+$2135 &  18.111 $\pm$  0.367 &  20087 &  18.50 $\pm$  0.50 &  6.75 &  5.71 &   49510566218944768 &  19.505 $\pm$  0.239 &   80.83 &  5.03 &  5.55 & 0  &  0  & 0.06  & 14  & 3.658  \\ 
04209$+$1352 &  21.976 $\pm$  0.882 &  20284 &  21.09 $\pm$  0.51 &  3.45 &  6.26 & 3310615565476268032 &  21.093 $\pm$  0.033 &    2.22 &  1.12 &  6.05 & 0  &  0  & 0.03  & 12  & 0.307  \\ 
04247$+$0442 &  16.700 $\pm$  0.033 &  20601 &  15.20 $\pm$  1.35 & -0.21 &  9.06 & 3283823387685219328 &  17.308 $\pm$  0.119 &   85.37 &  6.44 &  8.73 & 1  &  0  & 0.03  & 16  & 0.245  \\ 
04256$+$1556 &  22.366 $\pm$  0.728 &  20661 &  21.20 $\pm$  0.92 &  8.42 &  6.56 & 3312783561888068352 &  21.108 $\pm$  0.496 &  343.28 & 23.99 &  6.34 & 0  &  0  & 0.08  & 18  & 2.622  \\ 
04268$+$1052 &  21.758 $\pm$  1.356 &  20751 &  24.11 $\pm$  1.59 &  0.71 &  9.58 & 3306007129990778368 &  21.159 $\pm$  0.395 &  227.23 & 22.14 &  9.10 & 0  &  1  & 0.03  & 16  & 1.134  \\ 
\hline
\end{tabular}
\tablefoot{
Column~1. Identifier used in this paper. 
Column~2. Orbital parallax.
Column~3-6. \Hp\ data: HIP number, parallax, GOF, $H_{\rm p}$ magnitude.
Column~7-16. GDR3 data: Source identifier, parallax, GOF, RUWE, $G$ magnitude, Non-Single-Star (NSS)-flag, ipdfm ({\tt ipd\_frac\_multi\_peak}),
idpha ({\tt ipd\_gof\_harmonic\_amplitude}), Nv ({\tt visibility\_periods\_used}).
The last column is the significance of the modified Bp-Rp excess (C$^\star$), calculated according to \citet{Riello2021}.
}

\end{sidewaystable*}

\section{Results} 
\label{S-Res}

\subsection{Effects of pre-filtering in the NSS processing}
\label{SS-NSS}

Table~\ref{Tab:NSS} shows that only 21 of the 180 (186 unique objects minus 6 stars not in DR3)
known binaries have a non-zero NSS flag.
In other words, 90\% of the known binaries have not been flagged as such by the NSS pipeline(s).
Although this is the reality of the current release, it might be instructive to investigate how
these binaries were overlooked.
\citet{Halbwachs22} described the processing of astrometric binary stars.
Stars were selected to be brighter than $G$ = 19~mag and have 12 or more visibility periods.
The first criteria has no impact, and the second removes 23 objects.
There is a selection on keeping sources with RUWE $<1.4$, removing 93 objects (of the  130 that have a RUWE listed),
on {\tt ipd\_frac\_multi\_peak} $\le 2$, removing 64 objects,
on {\tt ipd\_gof\_harmonic\_amplitude} $< 0.1$, removing 57 objects,
and on the significance of the modified Bp-Rp excess, $\mid$$C^{\star}$$\mid$/$\sigma_{C^\star}$ $< 1.645$, removing 113 objects.

Only 30 objects meet all pre-selection criteria. Additional objects are likely to have been removed in the processing 
and the post-processing steps (see details in \citealt{Halbwachs22}).
Similar criteria must have been adopted in the processing of spectroscopic binaries, but the paper describing this
has not been published at the time of submission. As the aim of the paper is to use improved parallaxes from the
astrometric binary pipeline, the details of the processing in the spectroscopic binary pipeline are less relevant here.

\subsection{Comparing the orbital elements of the NSS processing with the literature}
\label{SS-compare}

Comparison of the the orbital elements in Table~\ref{Tab:NSS} with those in the literature reveals that a
number of them (03396$+$1823, 04179$-$3348, 12313$+$5507, 17038$-$3809, 17422$+$3804, 18339$+$5144, 23456$+$1309,
23485$+$2539, and  18339$+$5144) refer to a different component than listed in Table~\ref{Tab:Sam}, because period, eccentricity
or velocity amplitude do not match.
In all cases except one, the period in Table~\ref{Tab:NSS} is the shorter one, suggesting that the elements refer to some
inner orbit of the multiple system. Consultation of the  ORB6 suggests that none of these orbits was known.

In the other cases, the elements found by the NSS analysis are not more precise than known in the literature, which were therefore kept.
The median and 1.4826 $\cdot$ median absolute deviation (MAD; equivalent to 1$\sigma$ in a Gaussian distribution) were
calculated of ($x_1 - x_2)/\sqrt{\sigma_{x_1}^2 + \sigma_{x_2}^2}$, where $x$ represents period, eccentricity,
or the velocity amplitudes, from the literature and the NSS analysis, and which are expected to be zero and unity, respectively.

Although the sample is small the errors on the parameters in the SB2 solution appear to be underestimated, as is indeed suggested
at the end of section~6 in \cite{Babusiaux22}. When we take the formal errors, the width of the distribution is about 4-9$\sigma$.
When the errors are Scaled with $\sqrt{{\rm GOF}}$, this is reduced to 0.8-1.5$\sigma$.
For the non-SB2 solutions, the width is 0.6-1.6$\sigma$, suggesting that the error estimates are realistic.

\subsection{PZPO}
\label{SS-PZPO}

\begin{figure}
  \centering

\begin{minipage}{0.49\textwidth}
\resizebox{\hsize}{!}{\includegraphics{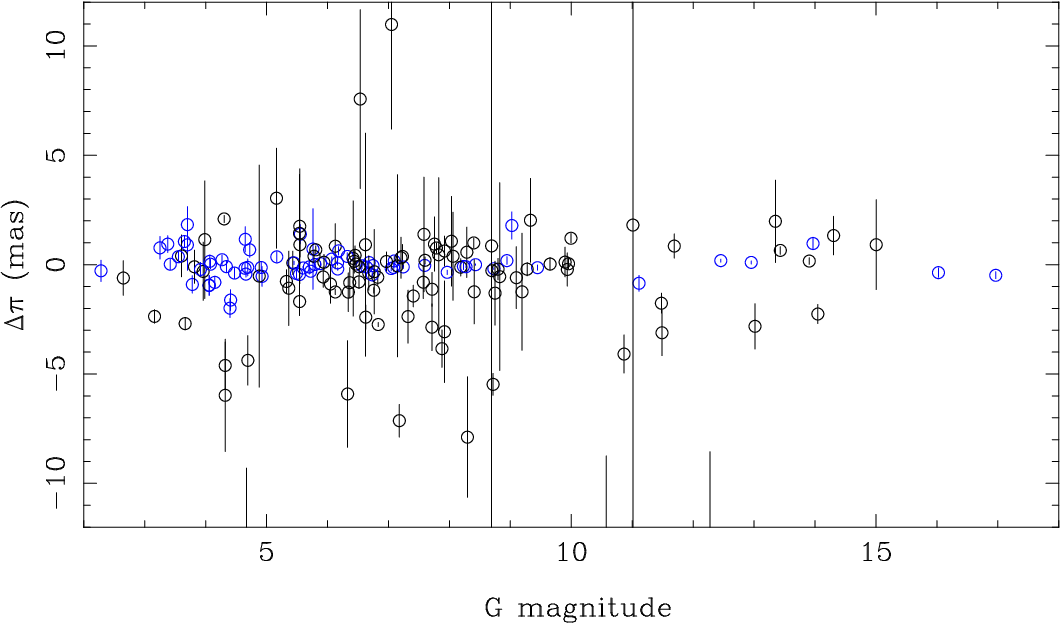}}
\end{minipage}

\caption{Parallax difference in the sense DR3 minus orbital parallax for the entire sample.
  The range in ordinate is $\pm$12~mas.
  The stars from the SS are plotted in blue (see text).
}
\label{Fig:PZPO1}
\end{figure}

\begin{figure}
  \centering

\begin{minipage}{0.49\textwidth}
\resizebox{\hsize}{!}{\includegraphics{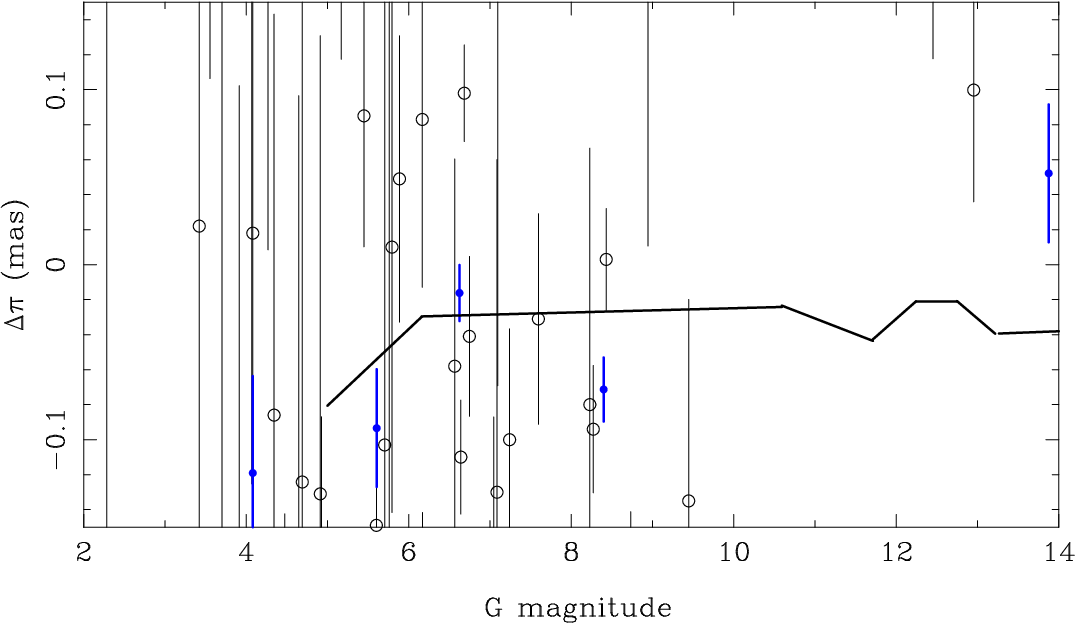}}
\end{minipage}

\caption{Zoom of Figure~\ref{Fig:PZPO1}, with the SS applied as described in the text.
  There are points outside the plot range.
  The blue points are the weighted averages plotted at the mean magnitude of the objects in the magnitude bins.
  The solid lines represent the magnitude dependence of the PZPO from G21 with an average spatial correction
  of $-0.013$~mas added.
}
\label{Fig:PZPO2}
\end{figure}

The main aim of the paper is to the investigate the PZPO.
Figure~\ref{Fig:PZPO1} shows the difference of DR3 parallax minus orbital parallax for the objects plotted against $G$ magnitude.
The range in ordinate is $\pm 12$~mas to display all points.
The \G\ parallax is the improved parallax from the NSS analysis (with its associated GOF parameter) for the eight stars in
Table~\ref{Tab:NSS} and from Table~\ref{Tab:Sam} otherwise. Added are the 37 WBS from Table~\ref{Tab:WBS} for a total of 170 determinations plotted.
When two  \G\ parallaxes were available (from the counterpart of the orbital parallax source and from a WBS), they were not averaged
as the PZPO depends on magnitude and possibly on colour.
The figure shows that some of the parallax errors (dominated by the error in orbital parallax) are large and
do not constrain any difference with the DR3 parallaxes.
Unless specified otherwise, a standard selection (SS) is applied from now on choosing objects where
the error in the orbital parallax should be smaller  than five times the error in the DR3 parallax,
the error in the orbital and DR3 parallaxes should be smaller than 2~mas,
the ratio of the absolute difference between orbital and DR3 parallax to the combined error bar should be lower than five, and
the GOF (either from the astrometric solution or the NSS solution) should be lower than 100, reducing the number of points to 68.
In Figure~\ref{Fig:PZPO1}, these points are plotted in blue, and Figure~\ref{Fig:PZPO2} shows a zoom with ordinates of $\pm$0.15 mas using this selection.
The data were also binned in $G$ mag, using five bins that started at $G$= 0, 5.0, 6.0, 7.4, and 9.2~mag (this last bin includes all fainter objects), 
where the weighted mean (and error on the mean) was calculated, and plotted at the mean $G$ magnitude of the objects in that bin. 
The first bin collects all of the brightest objects. Bright limits of 5.0 and 6.0~mag were used in G21 and L21/MA22, respectively, while 
7.4 and 9.2~mag are cardinal magnitudes used in MA22.
The weighted average over all objects results in an offset of $-41.7 \pm 10.7$ \muas\ (model 1 in Table~\ref{Tab:averages}, which also
includes the values per magnitude bin). 

Several corrections to the G(E)DR3 parallaxes have been proposed, and it is interesting to compare the corrected parallaxes to
the orbital ones. To guide the eye, the solid line in Figure~\ref{Fig:PZPO2} shows the magnitude correction from G21
with a constant spatial offset of $-0.013$~mas added (the average spatial correction of the sample following G21
at HEALPix level 0).
We recall that the HEALPix formalism \citep{HEALPix05} is a convenient way to divide the sky into equal-area pixels.
  At HEALPix level 0, there are 12 pixels, and this increases by a factor of four for every next higher level.
  The HEALPix formalism is used by the \G\ team and is encoded in
  the {\tt source\_id}\footnote{pixel number = {\tt source\_id}/(2$^{35} \cdot 4^{(12-{\rm level})})$ for a given HEALPix level.}.

Below, we apply different corrections on a star-by-star basis to the GDR3 parallaxes.
If the corrections work the weighted mean difference should be consistent with zero.
The first correction is the one by L21 (model 2). The number of sources is reduced to 30 as the correction is only defined
for stars fainter than magnitude 6. After correction, the difference with the orbital parallax is $-8.7 \pm 11.5$~\muas.
Model 3 shows the results for the MA22 correction, which is an extension of L21. The resulting difference is marginally
closer to zero. We note that the errors on the correction are identical to applying no correction. This is
related to the fact that the L21 and MA22 provided the offset without error bars.
Finally, the G21 formalism was tested. The correction depends on the chosen HEALPix level and consists
of a spatial correction defined at $G$= 20~mag (depending on the source coordinates (i.e. a certain pixel) at a
chosen HEALPix level) and an additive magnitude correction.
Following G21, only pixels with more than 40 QSOs have been considered.
Models 4-8 show the results for HEALPix levels 0-4, respectively. As the G21 correction is defined for $G>5,$ there are
more objects at the lower HEALPix levels, but then the number decreases with increasing HEALPix level
as the number of objects in pixels with insufficient QSOs increases.
HEALPix level 2 seems a good compromise between the sampling of the spatial correction and the loss of objects, and the
result is comparable to the results for the other correction methods. If the G21 correction is limited to $G>6$~mag, as it
is for the other two methods, the resulting difference is very close to zero (model 6a).

The SS is useful to obtain insight into the overall behaviour of the PZPO but now a stricter final selection (FS) is introduced.
As there are few objects fainter than tenth magnitude that nevertheless cover a wide range, and because there are very few extremely
bright objects, the sample was restricted in magnitude and two bins from 4-6.162 and 6.162-10.591~mag
(two breakpoints in Eq.~7 in \citealt{Gr21}) were considered.
As there is a general offset between orbital and \G\ parallaxes, the condition on the difference of the two was modified to
$\mid$$({\pi}_{\G}- (-0.04~{\rm mas}) ) - {\pi}_{\rm orb}$$\mid$/$\sigma_{c}$ $< 5.0$, where $\sigma_{c}$ is the combined error of the \G\
and the orbital parallax $\sqrt{  (k \cdot \sigma_{\pi_{\G}})^2 +  \sigma_{\pi_{\rm orb}}^2}$. 
The factor $k$ is the error inflation factor, which was set to unity in the SS.
Several papers have found that the error bars in the astrometric solution are underestimated \citep{Fabricius21, MA21, MaizA22}.
Here the formalism by \citet{ElBadry21} was used. There correction was derived for $G < 7$~mag, but it will be used for brighter magnitudes as well.
The effect is small in any case, $k= 1.10-1.15$ for $G < 10$~mag.
Finally, a more stringent cut on the GOF was imposed. The distribution of the GOF of QSOs was discussed in \citet{Gr21}, and in that paper
an interval from $-4$ to $+5$ was chosen for acceptable solutions.
Here the upper limit is slightly relaxed to $+8$, resulting in a sample of 20 objects. This is consistent with the other often-used selection
criterion of RUWE $<1.4$, which would result in 23 objects.

\begin{figure}
  \centering

\begin{minipage}{0.49\textwidth}
\resizebox{\hsize}{!}{\includegraphics{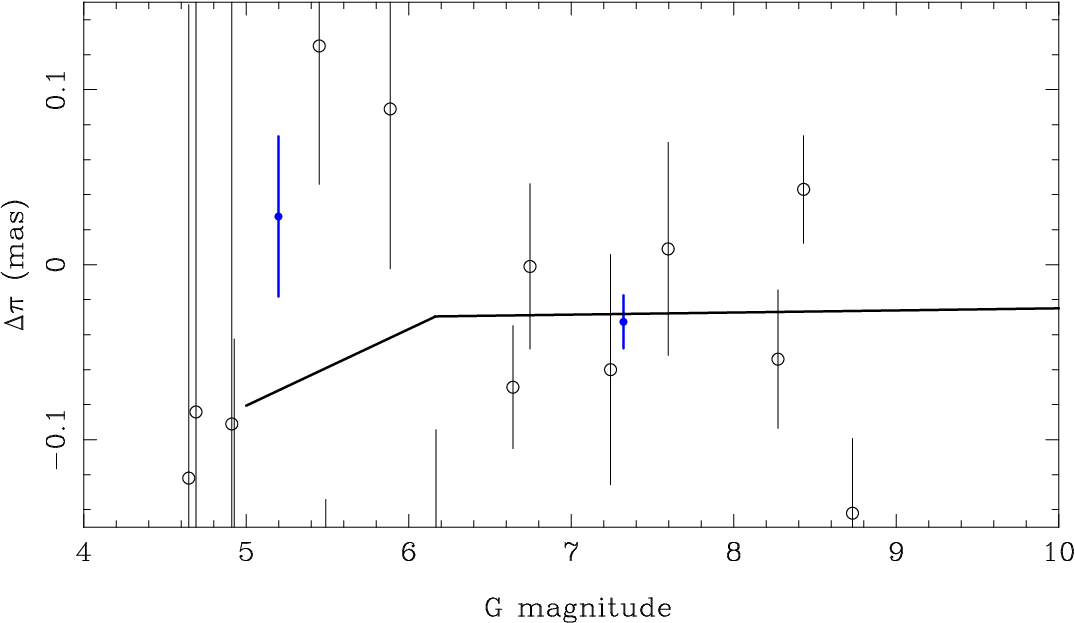}}
\resizebox{\hsize}{!}{\includegraphics{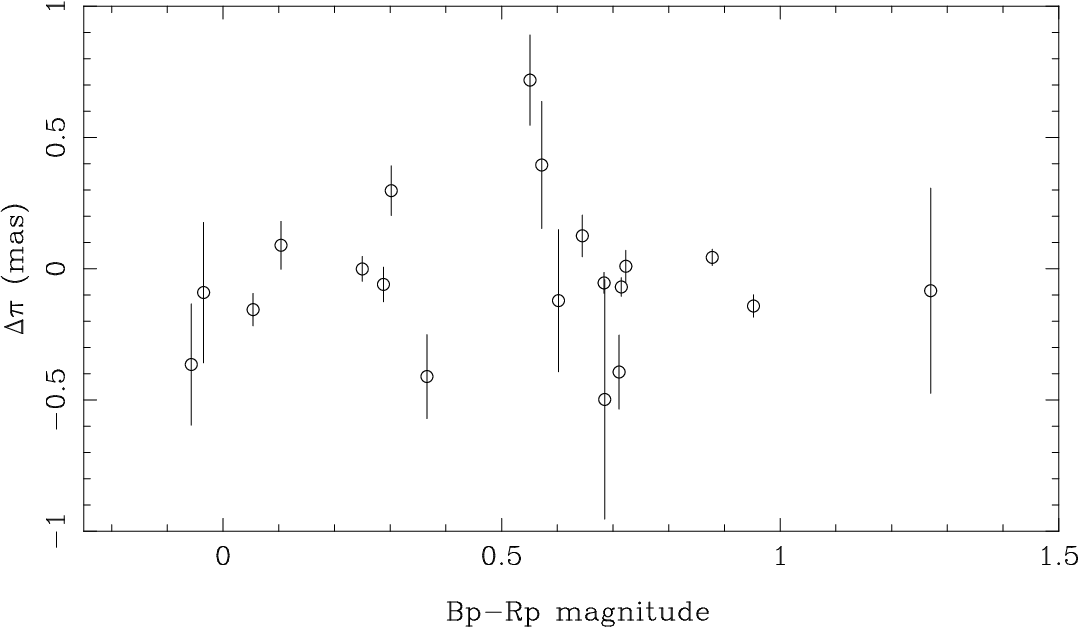}}
\resizebox{\hsize}{!}{\includegraphics{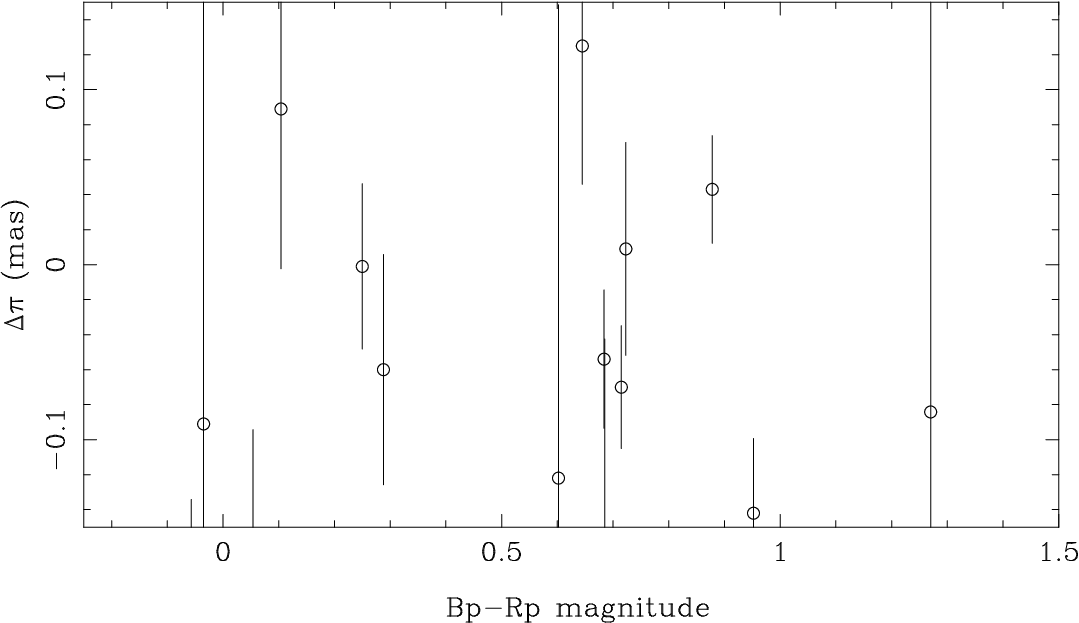}}
\end{minipage}

\caption{Parallax difference.
  Top panel: As Figure~\ref{Fig:PZPO2} for the final selection as described in the text.
  There are points outside the plot range.
  Middle panel: Offset plotted against Bp-Rp colour. The ordinate is chosen to be $\pm$ 1 mas to show all data points in the sample, and the 
  bottom panel has the same range in ordinate as in the top panel.
}
\label{Fig:PZPO3}
\end{figure}

Model 9 in Table~\ref{Tab:averages} gives the results, and they are displayed in Figure~\ref{Fig:PZPO3}. This figure also shows the offset as a function
of Bp-Rp colour with the range in ordinate chosen to show all objects, indicating that there is no evident dependence on colour.
Models 10-16 show the results after applying the various correction methods.
The results are very similar as before. All three proposed models bring the differences closer to zero, and consistent with zero
within the error bars.
When restricted to $G$ magnitudes fainter than 6, G21 again provides the correction closest to zero for HEALPix level 2.
Nevertheless, most of the results suggest that the corrections may be slightly underestimated (i.e. should be more negative), at least
on average over the 6-10 magnitude range.

\begin{table*}
  \centering

\caption{Parallax differences for different assumptions.}
\begin{tabular}{crrr|crrr|l} \hline \hline 
 Model   & $\Delta \pi$   & $G$   & N  &  Model   & $\Delta \pi$   & $G$   & N & Remarks  \\
         &      (\muas)   & (mag) &    &          &      (\muas)   & (mag) &   &            \\  \hline
 \multicolumn{4}{c}{\underline{standard selection (SS)}} & \multicolumn{4}{c}{\underline{final selection (FS)}} &    \\ 
 1     & $-41.7 \pm 10.7$ &  6.5 & 68 &  9     & $-26.7 \pm 14.3$ &  6.2 & 20 & all \\
       & $-119 \pm 55.4$  &  4.1 & 27 &        & $+78.9 \pm 41.1$ &  5.3 & 12 & G$<$5 (SS), 4-6.2 (FS) \\
       & $-93.4 \pm 33.7$ &  5.6 & 11 &        & $-41.1 \pm 15.3$ &  7.5 &  8 & G=5-6 (SS), 6.2-10.6 (FS) \\
       & $-16.2 \pm 16.0$ &  6.6 & 14 &        &&&& G=6-7.4  \\
       & $-71.4 \pm 18.3$ &  8.4 &  8 &        &&&& G=7.4-9.2 \\
       & $+52.2 \pm 39.4$ & 13.9 &  8 &        &&&& G$>$9.2  \\
 2     &  $-8.7 \pm 11.5$ &  9.0 & 30 &   10    & $-7.7 \pm 15.1$ &  7.3 & 9 &  all, L21 correction \\
       &  $+7.8 \pm 16.0$ &  6.6 & 14 &     &&&& G=6-7.4 \\
       & $-48.3 \pm 18.3$ &  8.4 &  8 &     &&&& G=7.4-9.2 \\
       & $+75.6 \pm 39.4$ & 13.9 &  8 &     &&&& G$>$9.2  \\
 3     &  $-7.6 \pm 11.5$ &  9.0 & 30 &   11    & $-8.8 \pm 15.1$ &  7.3 & 9 &  all, MA22 correction \\
       & $+15.5 \pm 16.0$ &  6.6 & 14 &     &&&& G=6-7.4  \\
       & $-57.0 \pm 18.3$ &  8.4 &  8 &     &&&& G=7.4-9.2 \\
       & $+82.6 \pm 39.4$ & 13.9 &  8 &     &&&& G$>$9.2  \\
 4     &  $-5.6 \pm 10.9$ &  8.1 & 41 &  12 & $+10.0 \pm 14.5$ & 6.6 & 15 & all, G21, HEALPIx level 0 correction \\
       & $-37.5 \pm 34.2$ &  5.6 & 11 &     &&&&  G=5-6 (SS) \\
       & $+14.1 \pm 16.1$ &  6.6 & 14 &    &&&&  G=6-7.4   \\
       & $-40.4 \pm 18.3$ &  8.4 &  8 &     &&&& G=7.4-9.2  \\
       & $+79.1 \pm 39.4$ & 13.9 &  8 &     &&&&  G$>$9.2  \\
 5     &  $-7.1 \pm 10.9$ &  8.1 & 41 & 13 & $+9.5 \pm 14.5$ &  6.6 & 15 &  all, G21 level 1 correction \\ 
       & $-41.0 \pm 34.2$ &  5.6 & 11 &      &&&&  G=5-6 (SS) \\
       & $+13.4 \pm 16.1$ &  6.6 & 14 &     &&&&  G=6-7.4  \\
       & $-42.2 \pm 18.4$ &  8.4 &  8 &     &&&&  G=7.4-9.2 \\
       & $+76.5 \pm 39.4$ & 13.9 &  8 &     &&&&  G$>$9.2  \\
 6     &  $-8.3 \pm 11.2$ &  8.2 & 39 & 14 & $+10.8 \pm 15.0$ &  6.7 & 14 &  all, G21 level 2 correction \\
       & $-58.3 \pm 37.8$ &  5.6 & 10 &     &&&&   G=5-6 (SS) \\
       & $+11.7 \pm 16.5$ &  6.7 & 13 &     &&&&  G=6-7.4  \\
       & $-40.4 \pm 18.5$ &  8.4 &  8 &     &&&&  G=7.4-9.2 \\
       & $+78.6 \pm 39.6$ & 13.9 &  8 &     &&&&  G$>$9.2  \\
 6a     &  $-3.5 \pm 11.7$ &  9.1 & 29 & 14a   & $+0.5 \pm 15.4$ &  7.3 &  9 &  G21, level 2 correction, $G>6$~mag \\
 7     &  $+5.6 \pm 12.4$ &  8.4 & 35 &  15  & $+11.4 \pm 17.1$ & 6.7 & 10 & all, G21, level 3 correction \\
       & $-45.8 \pm 46.0$ &  5.6 &  9 &     &&&&  G=5-6  \\
       & $+52.3 \pm 19.7$ &  6.6 & 10 &     &&&&  G=6-7.4  \\
       & $-44.9 \pm 18.8$ &  8.4 &  8 &     &&&&  G=7.4-9.2 \\
       & $+80.2 \pm 39.8$ & 13.9 &  8 &     &&&&  G$>$9.2  \\
 8     & $+18.2 \pm 13.6$ &  8.6 & 33 &  16  & $+24.9 \pm 19.8$ & 6.9 & 8 & all, G21, level 4 correction \\
       & $-65.2 \pm 47.3$ &  5.6 &  8 &     &&&&   G=5-6  \\
       & $+99.6 \pm 24.2$ &  6.6 &  9 &    &&&&   G=6-7.4  \\
       & $-38.3 \pm 19.5$ &  8.4 &  8 &     &&&&  G=7.4-9.2 \\
       & $+97.2 \pm 40.8$ & 13.9 &  8 &     &&&&  G$>$9.2  \\

\hline
\end{tabular} 
\tablefoot{
  Results are given for the SS (Cols.~1-4) and the FS (Cols.~5-8), and give
  the model number,
  the weighted mean parallax difference with error,
  the average $G$-mag of the objects in the bin, and the 
  the number of objects considered.
  Remarks and further details are given in Column~9.
}
\label{Tab:averages}
\end{table*}

\section{Conclusions}
\label{S-Dis}

The results from the NSS analysis specific to GDR3 show the potential but also the limitations of the current release.
Of the sample of 186 known binaries compiled from the literature to have an orbital parallax, only 8 have an astrometric
and 13 have an spectroscopic orbit determined from the NSS analysis. Most objects are eliminated at the pre-filtering stage of the NSS analysis.
The analysis of the parallax difference between \G\ and orbital parallax is therefore strongly influenced by the 
large GOF parameter in the main astrometric catalogue, limiting the number of useful objects.
The PZPO corrections proposed by L21, G21, and M22 give similar residuals.
After these corrections were applied the remaining parallax differences were formally consistent with zero within the error bar for all three recipes.
The current data and analysis therefore do not prefer a particular PZPO correction scheme over the other two.

Several improvements may be anticipated in the near future. The number of systems for which an orbital parallax will become available
will likely grow, or orbital elements of existing systems will improve.
For about one-third of the systems, separate astrometric and spectroscopic orbits exist that were used to obtain the orbital parallax.
For consistency, the existing data could be combined to obtain a single solution.

The situation might improve by DR4.
The orbital parallax for 18 out of 20 stars in the final selection sample is accurate to better then 5\%, and is even better than the \G\ value for
6 stars.
In the full sample, 148 stars have an orbital parallax determination better than 5\%.
Having NSS solutions available for a eight times larger sample would lead to a significantly more precise determination of the
PZPO at bright magnitudes.
Whether this is realistic would depend on how much the criteria in the pre-filtering and post-processing stages stages could be relaxed.
The impact of the former has been discussed, but some of the stars in the sample may also have been eliminated at the latter stage.
\citet{Halbwachs22} mentioned three criteria that have been applied as well.
The one on parallax accuracy\footnote{ ${\pi}/{\sigma_{\pi}} > 20~000 / P$ (days), using the orbital parallax to evaluate the inequality.} would
eliminate about 45\% of the sample. The selection on eccentricity accuracy\footnote{ ${\sigma_{\rm e}} < 0.182 \log P -0.244$.}
would eliminate 1\% of the sample, and the one on the significance of the
photocenter major axis\footnote{ $a_0/{\sigma_{\rm a_0}} > 158 /P$ (days), but using the major axis of the true orbit instead to evaluate the inequality.}
about 4\%.
Alternatively, with the planned release of the complete astrometric and spectroscopic time-series data with DR4, 
the community could combine \G\ data with literature data in order to obtain the best-determined orbit.

\begin{acknowledgements}
MG is grateful to the International Space Science Institute (ISSI) for support provided to the
SH0T ISSI International Team (\url{https://www.issibern.ch/teams/shot/}).
This work has made use of data from the European Space Agency (ESA) mission {\it Gaia} 
(\url{http://www.cosmos.esa.int/gaia}), processed by the {\it Gaia} Data Processing and Analysis Consortium 
(DPAC, \url{http://www.cosmos.esa.int/web/gaia/dpac/consortium}). 
Funding for the DPAC has been provided by national institutions, in particular
the institutions participating in the {\it Gaia} Multilateral Agreement.
This research has made use of the SIMBAD database and the VizieR catalogue access tool 
operated at CDS, Strasbourg, France.
\end{acknowledgements}

\bibliographystyle{aa.bst}
\bibliography{references.bib}

\begin{appendix}

\section{Additional tables}

\begin{table*}

\caption{Parameters for the WBS sample}
\begin{tabular}{rrrrrrrrrrrrrrrrrr} \hline \hline 
Name       &  ${\pi}_{\rm orb}$ &  Sep       & Source ID  &    $\pi$     & GOF & RUWE & $G$    & NSS & Flag  \\
           &      (mas)       &  (\arcsec) &            &    (mas)     &     &      &  (mag) &     & \\
\hline
01028$+$3148 & 8.859 $\pm$ 0.203 & 56.4 & 314142124302719744 &    8.495  $\pm$    0.051  &   1.876  &   1.083  &     16.027  &  0  & y \\
02057$-$2423 & 22.331 $\pm$ 1.457 & 56.6 & 5121561759597725440 &   21.028  $\pm$    0.019  &  -2.238  &   0.910  &      8.749  &  0  &  \\
02124$+$3018 & 11.825 $\pm$ 5.068 & 4.0 & 300312157810876160 &   12.738  $\pm$    0.550  & 139.299  &  17.831  &      6.622  &  0  &  \\
02128$-$0224 & 24.783 $\pm$ 2.616 & 16.8 & 2494066897240089984 &   26.168  $\pm$    0.027  &  -0.162  &   0.989  &      7.582  &  0  & y \\
02442$-$2530 & 22.253 $\pm$ 0.158 & 12.5 & 5076722404106857472 &   22.440  $\pm$    0.079  &  65.411  &   4.285  &      8.944  &  0  & y \\
03396$+$1823 & 26.378 $\pm$ 2.052 & 10.1 & 56861660442751872 &   27.289  $\pm$    0.045  &   5.265  &   1.280  &     15.005  &  0  &  \\
03400$+$6352 & 23.682 $\pm$ 1.927 & 46.2 & 488099359330834432 &   23.359  $\pm$    0.018  &  -1.755  &   0.951  &      6.768  &  0  &  \\
04179$-$3348 & 18.036 $\pm$ 0.112 & 5.5 & 4870527586936020864 &   19.005  $\pm$    0.257  &  42.199  &   4.799  &     13.968  &  0  & y \\
04247$+$0442 & 16.700 $\pm$ 0.033 & 7.1 & 3283823383389256064 &   16.800  $\pm$    0.055  &  30.926  &   2.686  &     12.953  &  0  &  \\
04560$+$3021 & 6.906 $\pm$ 0.353 & 9.2 & 156899557664509440 &    6.718  $\pm$    0.162  &   0.835  &   1.026  &     18.085  &  0  &  \\
08317$+$1924 & 77.062 $\pm$ 8.257 & 9.9 & 662732115407952768 &   60.248  $\pm$    0.076  &  27.886  &   2.051  &     12.278  &  0  &  \\
09194$-$7739 & 14.907 $\pm$ 0.299 & 9.1 & 5203285503955597184 &   14.702  $\pm$    0.010  &  -1.873  &   0.926  &      9.277  &  0  & y \\
09498$+$2111 & 3.400 $\pm$ 0.879 & 12.7 & 640100317815763712 &    4.729  $\pm$    0.023  &   0.499  &   1.017  &     14.306  &  0  &  \\
12351$+$1822 & 9.043 $\pm$ 0.362 & 20.1 & 3947649169267207296 &    8.919  $\pm$    0.127  &  -2.466  &   0.855  &      4.690  &  0  &  \\
13196$+$3507 & 72.115 $\pm$ 29.549 & 17.8 & 1473166433840437888 &   73.923  $\pm$    0.033  &  17.809  &   1.678  &     11.011  &  0  & y \\
13239$+$5456 & 40.497 $\pm$ 0.143 & 14.4 & 1563590510627624064 &   40.280  $\pm$    0.285  &  52.627  &   3.826  &      3.914  &  0  &  \\
14206$-$3753 & 13.050 $\pm$ 0.063 & 57.4 & 6116979078226372992 &   13.233  $\pm$    0.018  &   3.244  &   1.179  &     12.453  &  0  &  \\
14575$-$2125 & 170.963 $\pm$ 1.697 & 26.0 & 6232511606838403968 &  169.884  $\pm$    0.065  &   4.937  &   1.298  &      5.364  &  0  &  \\
15006$+$0836 & 26.385 $\pm$ 0.107 & 34.0 & 1161798072431380096 &   25.901  $\pm$    0.114  &   2.474  &   1.129  &     16.967  &  0  & y \\
15282$-$0921 & 50.720 $\pm$ 1.212 & 52.2 & 6317854118838346752 &   48.345  $\pm$    0.029  &   2.005  &   1.107  &      7.321  &  0  &  \\
16147$+$3352 & 43.980 $\pm$ 0.528 & 7.3 & 1328866562170960384 &   44.134  $\pm$    0.018  &   2.208  &   1.078  &      6.438  &  0  & y \\
16212$-$2536 & 6.239 $\pm$ 0.131 & 20.2 & 6048602103662751488 &    7.233  $\pm$    0.177  & 114.633  &   7.170  &      8.401  &  0  &  \\
16311$-$2405 & 10.466 $\pm$ 1.031 & 13.1 & 6050627782031927296 &    7.645  $\pm$    0.020  &   6.648  &   1.301  &     13.015  &  0  &  \\
17038$-$3809 & 15.436 $\pm$ 0.168 & 10.6 & 5976304983594844928 &   16.084  $\pm$    0.021  &   6.147  &   1.230  &     13.432  &  0  & y \\
17584$+$0428 & 21.666 $\pm$ 1.878 & 18.2 & 4472789731012862080 &   23.647  $\pm$    0.022  &   4.509  &   1.235  &     13.349  &  0  &  \\
18002$+$8000 & 22.998 $\pm$ 0.463 & 18.7 & 2294405721759384064 &   22.434  $\pm$    0.035  &   3.978  &   1.230  &      5.928  &  0  &  \\
18055$+$0230 & 194.423 $\pm$ 2.677 & 6.4 & 4468557611977674496 &  195.856  $\pm$    0.254  &  35.917  &   3.706  &      5.539  &  0  & y \\
18058$+$2127 & 24.576 $\pm$ 0.674 & 28.2 & 4576326312901650560 &   24.694  $\pm$    0.022  &  29.445  &   1.787  &      9.898  &  0  &  \\
18099$+$0307 & 20.785 $\pm$ 0.552 & 7.2 & 4469921487444001792 &   21.637  $\pm$    0.030  &   7.282  &   1.310  &     11.690  &  0  & y \\
18339$+$5144 & 60.424 $\pm$ 0.150 & 17.0 & 2145277550935526784 &   60.590  $\pm$    0.021  &   5.907  &   1.238  &     13.907  &  0  &  \\
18413$+$3018 & 8.379 $\pm$ 0.250 & 14.1 & 4587744672430944512 &    8.103  $\pm$    0.013  &  -4.908  &   0.833  &      8.699  &  0  &  \\
18501$+$3322 & 3.197 $\pm$ 0.931 & 45.8 & 2090687726329643392 &    3.513  $\pm$    0.090  &  44.763  &   2.862  &      7.205  &  0  &  \\
19091$+$3436 & 28.251 $\pm$ 0.853 & 16.0 & 2044341077844180736 &   24.413  $\pm$    0.015  &  -3.163  &   0.891  &      7.879  &  0  &  \\
19196$+$3720 & 41.081 $\pm$ 0.070 & 34.6 & 2051069745406938240 &   39.318  $\pm$    0.450  & 337.307  &  26.510  &     11.480  &  0  & y \\
19394$+$3009 & 14.766 $\pm$ 0.412 & 4.8 & 2032457178231864704 &   12.511  $\pm$    0.132  &   5.112  &   1.340  &     14.046  &  0  & y \\
21232$-$8703 & 14.556 $\pm$ 4.288 & 17.8 & 6342009495947730432 &   14.010  $\pm$    0.013  &  -1.170  &   0.953  &      8.826  &  0  &  \\
22038$+$6438 & 38.094 $\pm$ 2.432 & 8.1 & 2218144866573662848 &   32.184  $\pm$    0.018  &  -1.464  &   0.926  &      6.330  &  0  & y \\
\hline
\end{tabular} 
\tablefoot{
  Column~1-2: Name and orbital parallax of the main source.
  Column~3-4: Source Id of the WBS candidate and the separation between the two sources.
  Column~4-9: parallax (with error), GOF, RUWE, $G$-mag, and NSS-flag.
  Column~10: Flag indicating if the star is in \citet{ElBadry21}.
}
\label{Tab:WBS}
  \end{table*}



\begin{sidewaystable*}

    \small
\setlength{\tabcolsep}{1.3mm}
\caption{Parameters from the NSS analysis}
\begin{tabular}{rrrrrrrrrrrrrrrrrr} \hline \hline 
Name       &  Source ID          & NSS     & $\pi$    & $\sigma_{\pi}$  & Period & $\sigma_{\rm P}$ & $e$  & $\sigma_{\rm e}$  & GOF & Eff & Sign & $K_1$  & $\sigma_{\rm K_1}$ & $K_2$ & $\sigma_{\rm K_2}$ & $a_{\rm 0}$ &  $\sigma_{\rm a_0}$  \\
           &                     & model   &  (mas)   &     (mas)   &       (d)  &  (d)           &      &               &        &     &      &  (\ks) &    (\ks)       &   (\ks)  &    (\ks)      &  (mas) & (mas)    \\  \hline
\hline
00490$+$1656 & 2781872793183604096 & SB2 &     &     &   13.82738  &  0.00027  & 0.2464  & 0.0014  &   8.66  & 0.45  & 519.1  &  60.315  &  0.116  &  57.883  &  0.112  &       &     \\
01096$-$4616 & 4933562966514402176 & SB2 &     &     &   24.60049  &  0.00075  & 0.1896  & 0.0020  &  20.15  & 0.48  & 286.3  &  48.734  &  0.170  &  51.756  &  0.172  &       &     \\
03147$-$3533 & 5047189109469319680 & SB2 &     &     &   75.63557  &  0.00343  & 0.0382  & 0.0020  &  47.40  & 0.32  & 406.3  &  39.852  &  0.098  &  39.858  &  0.082  &       &     \\
03396$+$1823 & 56861694804053120 & SB2 &     &     &    6.81420  &  0.00007  & 0.4198  & 0.0035  &  36.56  & 0.40  & 176.8  &  69.329  &  0.392  &  74.454  &  0.379  &       &     \\
03566$+$5042 & 250437313946591360 & SB2 &     &     &   30.43678  &  0.00089  & 0.6488  & 0.0018  &  37.53  & 0.43  & 267.4  &  52.180  &  0.195  &  51.924  &  0.190  &       &     \\
04179$-$3348 & 4870527586937201408 & SB2 &     &     &    0.97019  &  0.00001  & 0.4518  & 0.0094  &  35.52  & 0.66  &  73.2  &  50.605  &  0.691  &  43.662  &  0.843  &       &     \\
04247$+$0442 & 3283823387685219328 & OTS          &   16.518  & 0.024  &  156.34427  &  0.15026  & 0.8978  & 0.0663  &   0.71  & 0.00  &  11.0  &      &      &      &      &   3.430  &   0.313  \\
04469$-$6036 & 4677731006145097984 & SB2 &     &     &   14.90132  &  0.00026  & 0.1965  & 0.0026  &  12.04  & 0.47  & 281.3  &  56.339  &  0.200  &  56.132  &  0.201  &       &     \\
04506$+$1505 & 3404812685132622592 & FDTSB1 &     &     &      &      &      &      &   1.53  &      &      &      &      &      &      &       &     \\
12313$+$5507 & 1571145907856592768 & OTS          &   39.605  & 0.019  & 1093.35755  &  5.22279  & 0.5371  & 0.0027  &   0.28  & 0.00  & 110.2  &      &      &      &      &  26.812  &   0.243  \\
14104$+$2506 & 1257529526205454336 & SB2 &     &     &    9.60584  &  0.00005  & 0.1996  & 0.0010  &  43.06  & 0.59  & 617.5  &  67.241  &  0.109  &  69.392  &  0.105  &       &     \\
16057$-$2027 & 6244076338858859776 & OTS          &   54.476  & 0.025  &  105.82585  &  0.05186  & 0.1731  & 0.0166  &   1.00  & 0.00  & 173.3  &      &      &      &      &   4.734  &   0.027  \\
17038$-$3809 & 5976304987919824384 & OTS          &   16.114  & 0.030  &    7.49295  &  0.00264  & 0.2412  & 0.1685  &   0.50  & 0.00  &   8.5  &      &      &      &      &   0.310  &   0.037  \\
17422$+$3804 & 1342735149009387136 & OTS          &   25.197  & 0.022  & 1236.58598  & 39.37131  & 0.0607  & 0.0222  &   0.70  & 0.00  &  34.7  &      &      &      &      &  12.454  &   0.359  \\
17422$+$3804 & 1342735149009387136 & SDTSB1 &     &     &      &      &      &      &   1.49  &      &      &      &      &      &      &       &     \\
18339$+$5144 & 2145277550935525760 & OTS          &   60.620  & 0.023  &   32.26596  &  0.03200  & 0.1077  & 0.0767  &   0.86  & 0.00  &  17.4  &      &      &      &      &   1.454  &   0.083  \\
18339$+$5144 & 2145277550935525760 & SB2 &     &     &    5.97735  &  0.00013  & 0.3555  & 0.0061  &  33.34  & 0.64  & 140.5  &  34.284  &  0.244  &  30.652  &  0.223  &       &     \\
19264$+$4928 & 2129771310248902016 & ASSB1 &   40.881  & 0.018  &  166.76852  &  0.08022  & 0.1342  & 0.0077  &   3.07  & 0.00  & 217.2  &      &      &      &      &   4.271  &   0.020  \\
19399$-$3926 & 6690041553622067328 & SB1 &     &     &   11.41470  &  0.00042  & 0.0867  & 0.0080  &   2.35  & 0.32  & 112.4  &  42.009  &  0.374  &      &      &       &     \\
20329$+$4154 & 2067948245320365184 & Orb &   46.052  & 0.021  &   57.34178  &  0.02049  & 0.3118  & 0.0185  &   5.08  & 0.26  & 184.9  &      &      &      &      &   4.406  &   0.024  \\
23456$+$1309 & 2770028132375790976 & SB2 &     &     &   19.89153  &  0.00086  & 0.3655  & 0.0026  &  40.21  & 0.57  & 164.4  &  39.733  &  0.242  &  29.740  &  0.241  &       &     \\
23485$+$2539 & 2852594583674129152 & SB1 &     &     &  977.91854  & 104.31029  & 0.5867  & 0.1500  &   1.58  & 0.31  &   8.9  &   1.188  &  0.134  &      &      &       &     \\
23571$+$5542 & 1994714276926012416 & SB2 &     &     &   12.15694  &  0.00011  & 0.3156  & 0.0016  &  10.97  & 0.46  & 347.7  &  71.700  &  0.206  &  72.416  &  0.207  &       &     \\
\hline
\end{tabular} 
\tablefoot{
  Column~1: Name.
  Column~2: Source Id.
  Column~3: NSS model. OTS stands for OrbitalTargetedSearch, ASSB1 stands for AstroSpectroSB1, Orb stands for Orbital, FDTSB1 stands for FirstDegreeTrendSB1,
  SDTSB1 stands for SecondDegreeTrendSB1.
  Column~4-18: parallax, period, eccentricity (with errors), GOF, {\tt efficiency}, {\tt significance}, velocity amplitude
  of the two components (with errors), photocenter semi-major axis with error.
}
\label{Tab:NSS}
  \end{sidewaystable*}

\end{appendix}

\end{document}